\documentclass[sigconf]{acmart}

\usepackage{placeins}
\setcopyright{none}
\renewcommand\footnotetextcopyrightpermission[1]{}
\begin{document}

\title{COVID, BLM, and the polarization of US politicians on Twitter}



\author{Anmol Panda}
\affiliation{%
  \institution{Microsoft Research India}
  \city{Bengaluru}
  \country{India}
}

\author{Divya Siddarth}
\affiliation{%
  \institution{Microsoft Research India}
  \city{Bengaluru}
  \country{India}
}

\author{Joyojeet Pal}
\affiliation{%
  \institution{Microsoft Research India}
  \city{Bengaluru}
  \country{India}
}





\begin{abstract}
 We mapped the tweets of 520 US Congress members, focusing on analyzing their engagement with two broad topics: first, the COVID-19 pandemic, and second, the recent wave of anti-racist protest. We find that, in discussing COVID-19, Democrats frame the issue in terms of public health, while Republicans are more likely to focus on small businesses and the economy. When looking at the discourse around anti-Black violence, we find that Democrats are far more likely to name police brutality as a specific concern. In contrast, Republicans not only discuss the issue far less, but also keep their terms more general, as well as criticizing perceived protest violence.
  
  \end{abstract}





\maketitle
\section{Introduction}
For years, politicians have been using social media to communicate directly with voters, and researchers have studied this phenomenon\cite{katz2013social, enli2017twitter, metzgar2009social, kreiss2012acting, talbot2008obama, hughes2010obama}. During this time, the study of politicians' social media use has covered a wide range of topics, with focus on partisan messaging \cite{hemphill2013framing}, engagement with constituents, polarization, \cite{khondker2011role, alsayyad2015virtual}, volunteer mobilization \cite{boulianne2015social}, targeting of opponents, and propaganda as observed in Germany \cite{arzheimer2015afd}, Austria \cite{wodak2017right}, Italy \cite{caiani2009dark}, France \cite{benveniste2016far} and India \cite{mohan2015locating}. 

In the present article, we contribute to this space with an analysis of tweets shared by the official handles of 519 members of the US Congress (ninety-seven Senators and four hundred and twenty two members  of the House of Representatives who have active and verifiable accounts on Twitter). The period of interest was from Jan 1, 2020 to June 29, 2020. This period was chosen specifically to study the politicians' framing of the COVID-19 pandemic, and accompanying public health and economic crises, as well to capture discourse around the recent wave of anti-racist protest across the country. 

In this study, we intend to answer the following questions:
\begin{enumerate}
    \item How did the two parties frame their discussion around three key topics: Covid-19, the George Floyd murder, and the Black Lives Matter movement?
    \item What does this framing say about the willingness to address or avoid specific issues?
\end{enumerate}

\section{Data and Methodology}

The primary list of politicians was hand-drawn, seeding from a list of members of Congress in late 2019 and searching for their handles on Twitter. It was complemented with more recent data and metadata from Github \footnote{Link: https://github.com/unitedstates/congress-legislators}. In all, we collected 310,686 tweets of 520 legislators of the US Congress using Twitter's public Search API. With over 10 million followers, the  @SenSanders handle of Senator Bernie Sanders (I-VT) was the most followed account in Congress, while @GuamCongressman, the official handle of Michael F.Q. San Nicolas, member of the House from the US territory of Guam, was the least followed with 628 followers at the time of the study. The accounts of @Jim\_Jordan (R-OH) and @RepAdamSchiff (D-CA) were the most retweeted for the two parties in the study period, with median retweet rates of 4485 and 11194 respectively. 

We labelled  both original and retweeted tweets for each of our three issues of interest: Covid19, George Floyd, and Black Lives Matter. We use the following lists of string literals to match for topic relevance:
\begin{itemize}
    \item Covid19: [`ncov', `covid', `corona', `virus']
    \item George Floyd: [`georgefloyd', `george floyd']
    \item Black Lives Matter: ['blm', 'blacklivesmatter', 'black lives matter', 'black lives']
\end{itemize}

We use two approaches for this labelling. In the first approach, we consider occurrence of the aforementioned terms in the tweet text of all tweets. For retweeted tweets, we extract the complete text of the original tweet. We call this the {\tt InText} category. For the second approach, we match the terms only in hashtags of tweets. We call this the {\tt InHashtag} category. In both cases, the matching is caseless and considers substrings i.e. `covid' will match to \#Covid19, \#CovidPandemic, etc. in the {\tt InHashtag} category, while it will match additionally to terms like `Covid19', `CovidCrisis', etc. in the {\tt InText} category.  
\section{Results}

In general, Democrats tweeted more on a weekly basis. When considering {\tt InText} tweets, they tweeted more across categories, with a greater number of mentions of Covid19, George Floyd and BLM than Republicans. The difference in Covid19 tweet counts was small, but significant. On the other hand, the difference in BLM and George Floyd-related tweets was substantial. When considering {\tt InHashtag} tweets, difference in Covid19 hashtag use by the two parties was not significant. For BLM and George Floyd, it was significant. We controlled for the user that posted the tweet, the week and the number of total tweets they posted in that week. 

\begin{figure*}
    \centering
    \includegraphics[width=\textwidth]{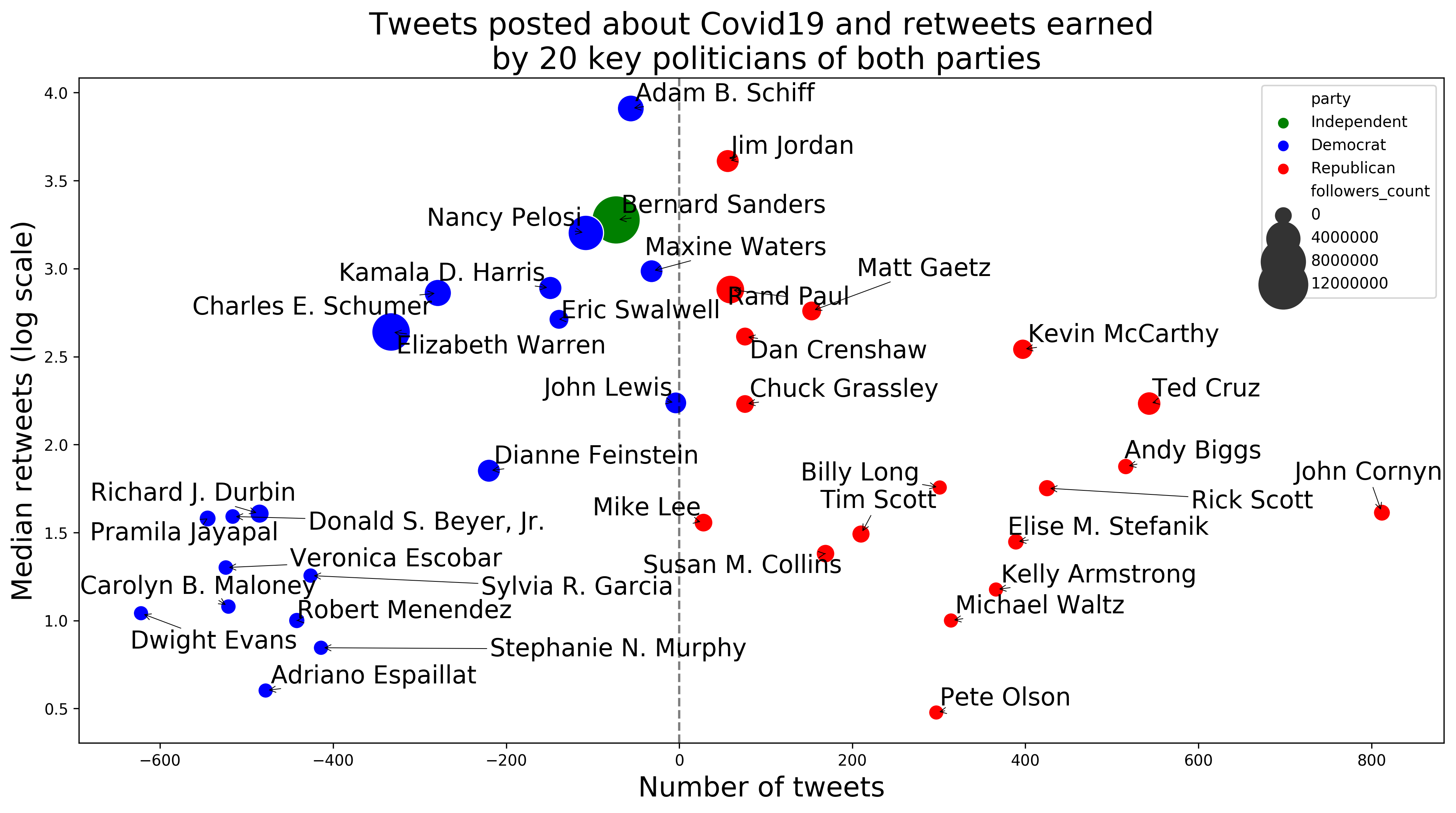}
    \caption{Tweet and retweet data of key members of US Congress on Covid19 
    ({\tt InText} category) }
    \label{fig:tweet_covid}
\end{figure*}

\begin{figure*}
    \centering
    \includegraphics[width=\textwidth]{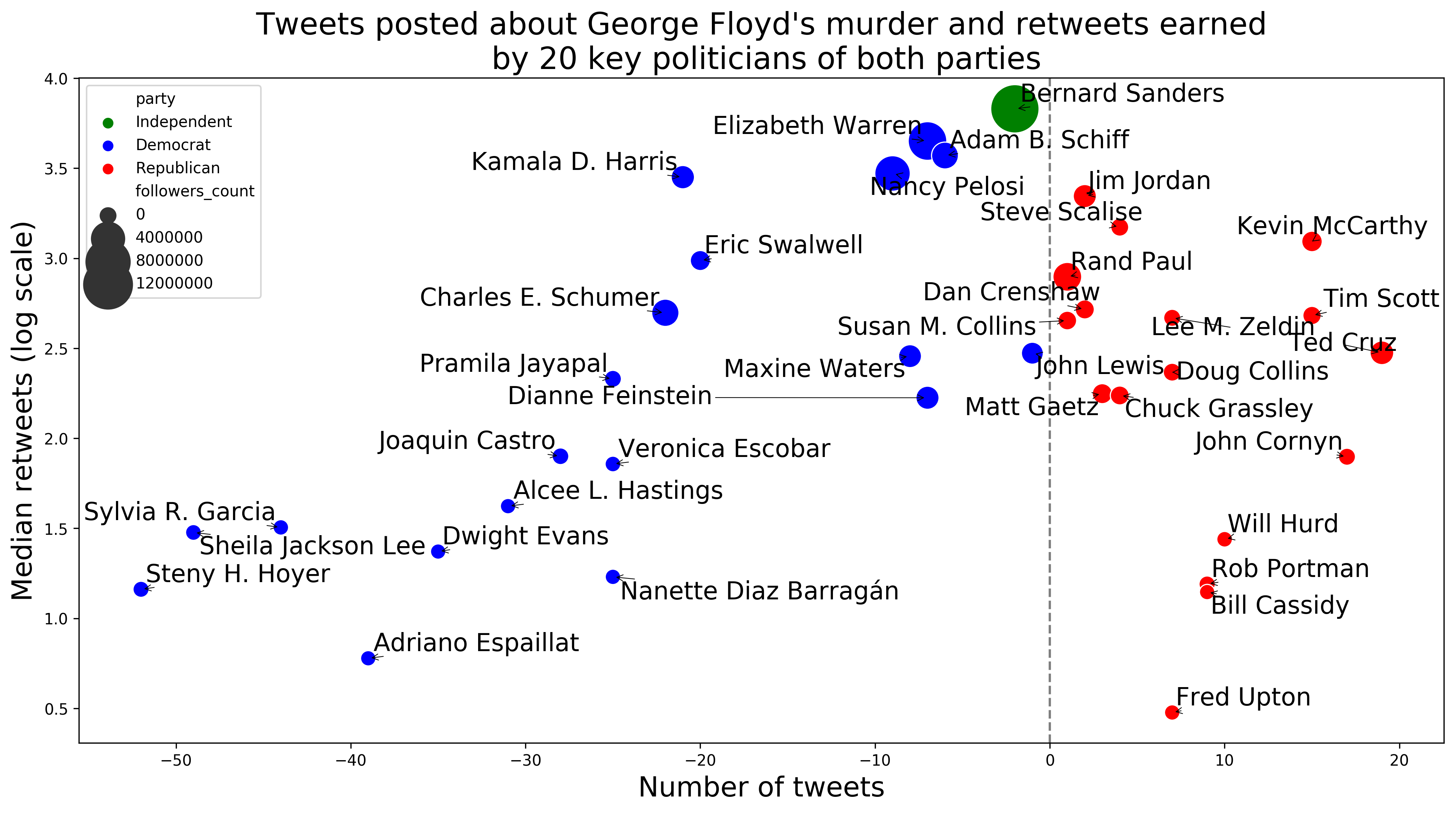}
    \caption{Tweet and retweet data of key members of US Congress on George Floyd's murder ({\tt InText} category) }
    \label{fig:tweet_george_floyd}
\end{figure*}

\begin{figure*}
    \centering
    \includegraphics[width=\textwidth]{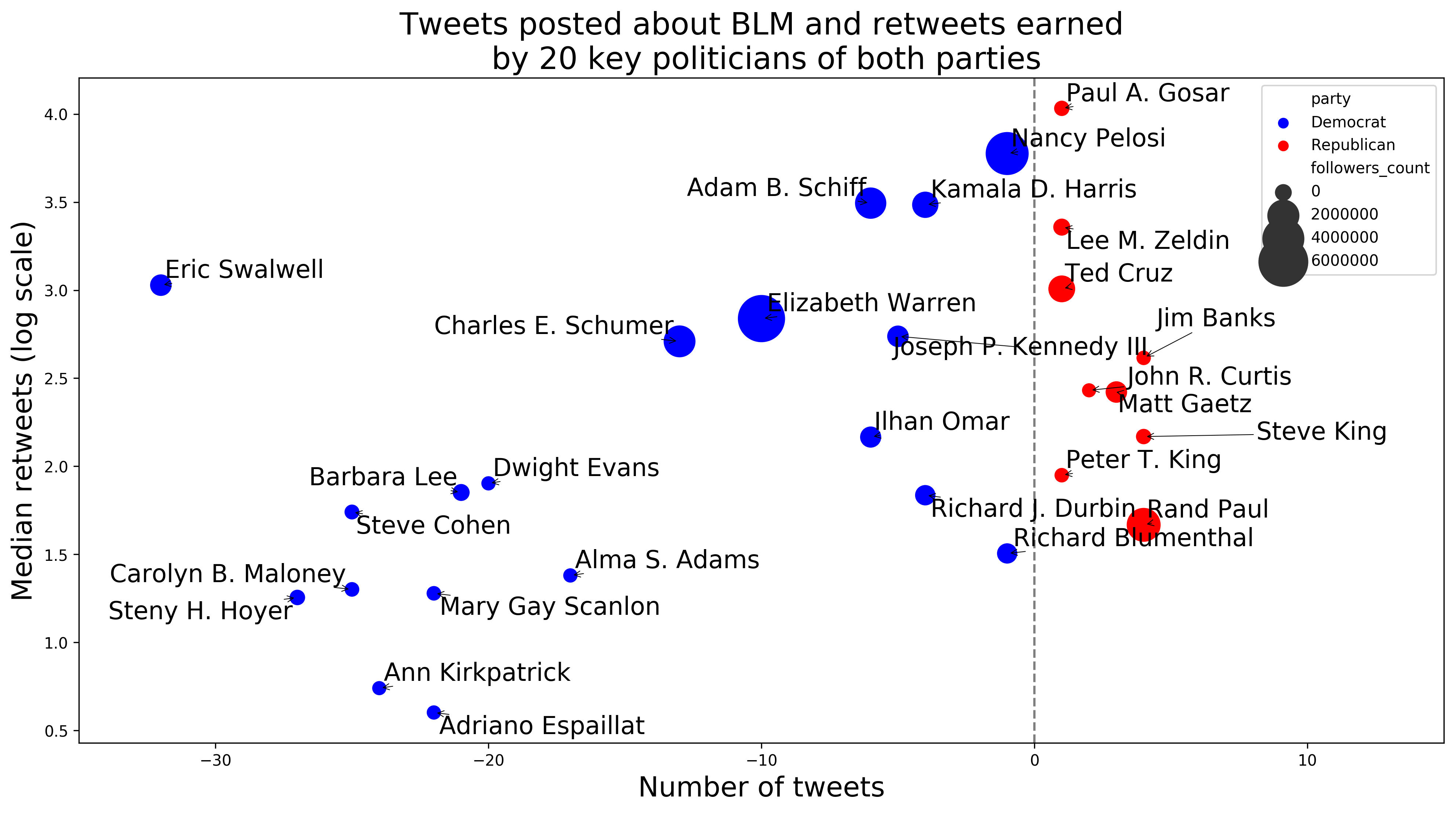}
    \caption{Tweet and retweet data of key members of US Congress on BLM 
    ({\tt InText} category) }
    \label{fig:tweet_blm}
\end{figure*}

We plotted the partisan divide in framing by the US Congress on the three key topics using two methods - by estimating how many and which politicians posted about those topics, and by analysing co-occurring terms in tweets about those topics. For the former, we plot the number of tweets posted by politicians and the median retweets they received in three interactive visualizations\footnote{Link to interactive visualizations: https://github.com/anmolpanda/US-Congress-Analysis/tree/master/visualizations/polarization-paper}. In these visualizations, the x-axis plots the number of tweets (for {\tt InText} category) posted by a member of US Congress, y-axis plots the median retweets they received, and the size of the bubble corresponds to the number of their Twitter followers. Figures \ref{fig:tweet_covid}, \ref{fig:tweet_george_floyd}, and \ref{fig:tweet_blm} show excerpts from these with the top ten most followed and top ten most active \footnote{Activity quantified by the number of tweets posted about the given topic during the study period} politicians from both parties represented for each topic. We observe that, with regards to Covid19, 274 Democrats posted 43491 tweets, the 2 independent Senators posted 142 tweets, and 240 Republicans posted 27155 tweets. Looking at tweets surrounding George Floyd's murder, 266 Democrats put out 2118 tweets, independent Senators posted 4 tweets, and 157 Republicans posted 464 tweets. On BLM, 191 Democrats tweets 1036 times while 9 Republicans posted 21 tweets We verified every BLM tweet to remove false positives, especially tweets about the Bureau of Land Management (@BLMNational). Prior to Floyd's murder (May25), Democrats posted 56 tweets about the Bureau and 21 about Black Lives Matter, while Republicans posted 59 tweets regarding the former and none about the latter. After the incident, Democrats posted 1015 tweets about the movement and only 11 about the Bureau, while the GOP posted 21 and 17 tweets respectively.  

\begin{figure*}
    \centering
    \includegraphics[width=\textwidth]{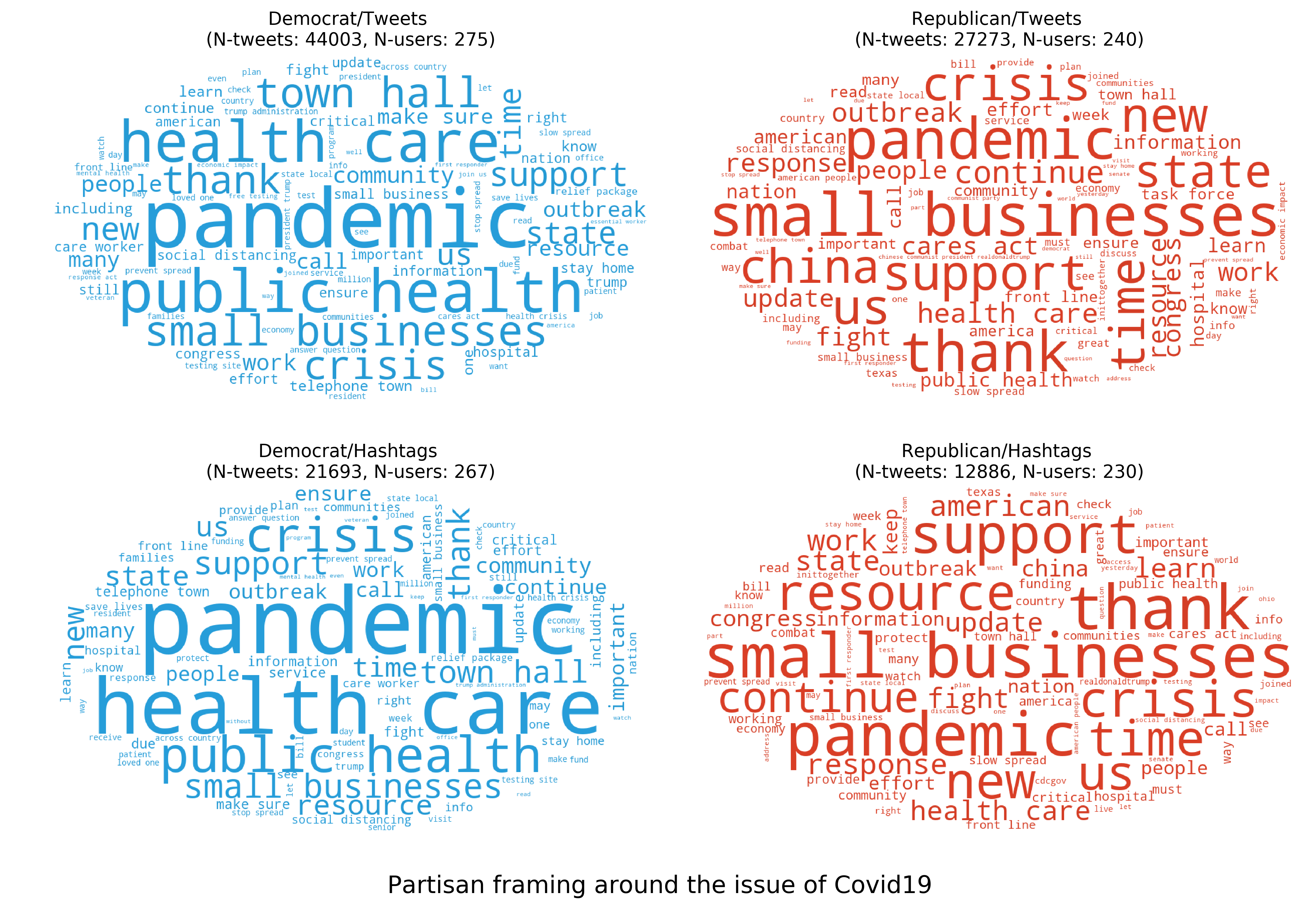}
    \caption{Word clouds of tweets relating to Covid19 by the two main parties (top) and word clouds of tweets with hashtags that relate to Covid19 (bottom) }
    \label{fig:covid}
\end{figure*}

\begin{figure*}
    \centering
    \includegraphics[width=\textwidth]{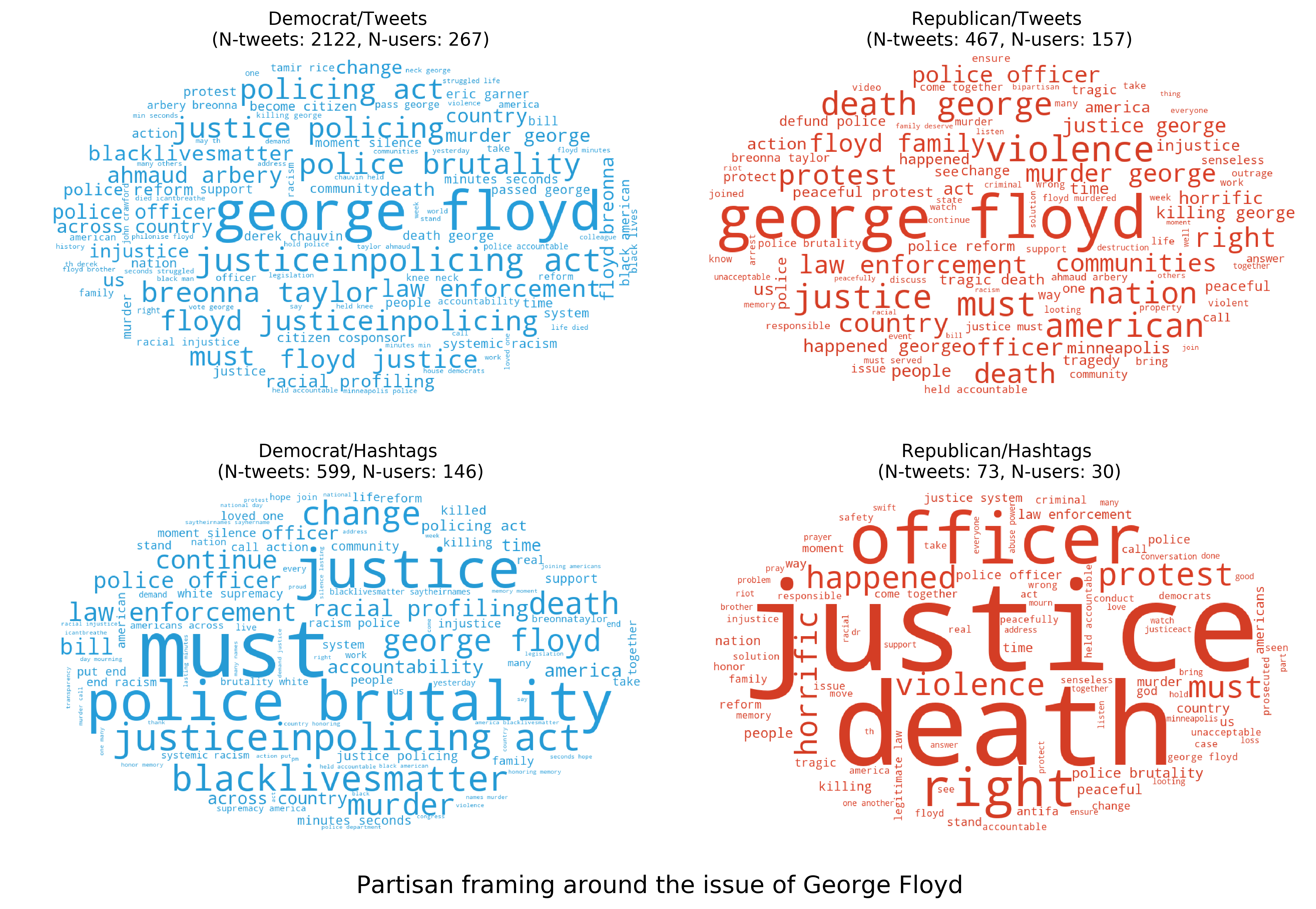}
    \caption{Word clouds of tweets relating to George Floyd by the two main parties (top) and word clouds of tweets with hashtags that relate to George Floyd (bottom) }
    \label{fig:george_floyd}
\end{figure*}

\begin{figure*}
    \centering
    \includegraphics[width=\textwidth]{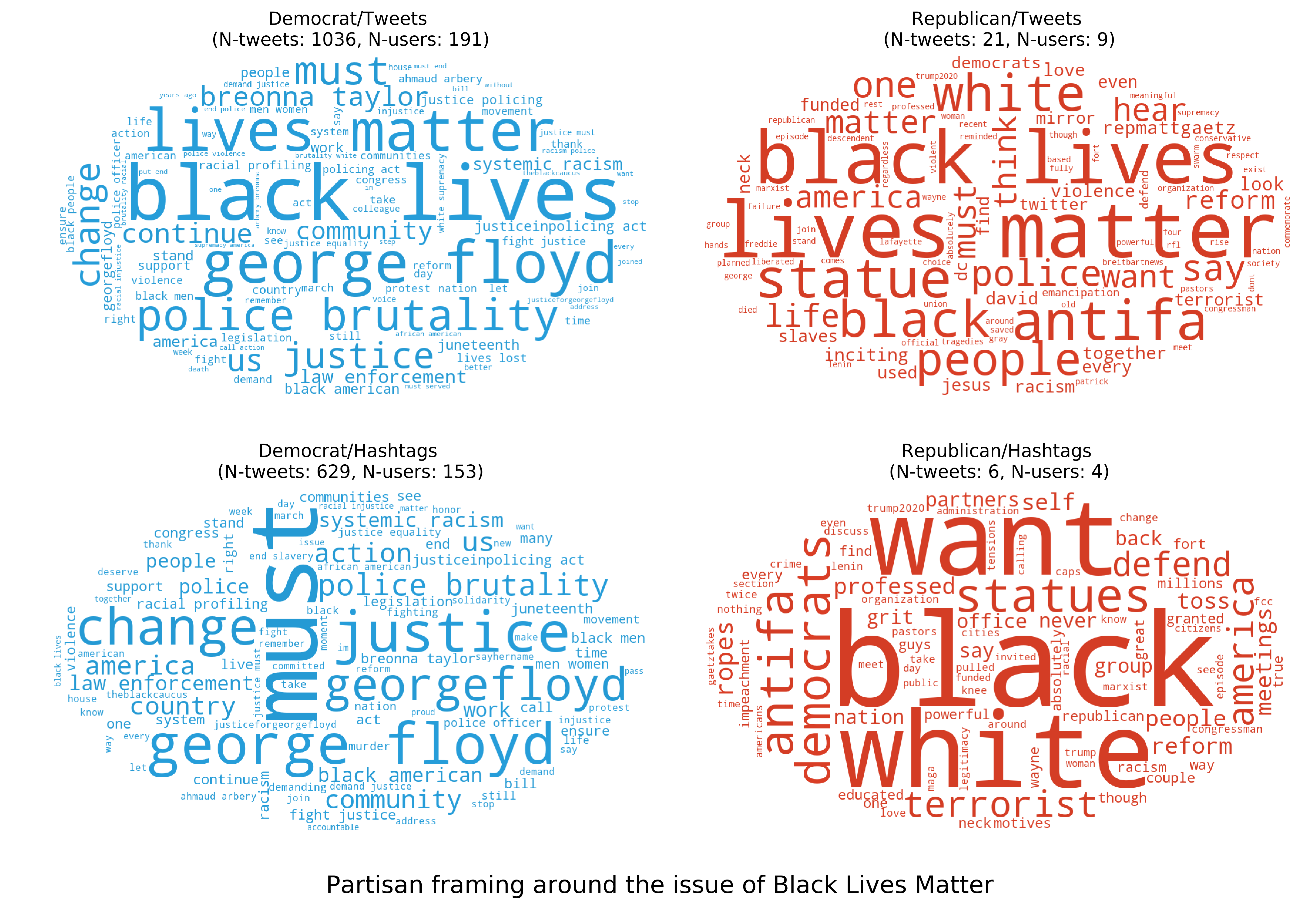}
    \caption{Word clouds of tweets relating to Black Lives Matter by the two main parties (top) and word clouds of tweets with hashtags that relate to Black Lives Matter (bottom) }
    \label{fig:blm}
\end{figure*}

Secondly, we plot the word clouds of terms in tweets about key topics. Figures \ref{fig:covid}, \ref{fig:george_floyd}, and \ref{fig:blm} display the word clouds of tweets of the two parties for both the {\tt InText} and {\tt InHashtag} categories. The results indicate the following:

On Covid19,
    \begin{itemize}
        \item The Democratic party's framing of the Covid crisis centered around healthcare, with terms like `public health', `pandemic' and `health care' emerging as the most frequent phrases in both the {\tt InText} and {\tt InHashtag} categories.
        \item Republicans, on the other hand, focused more on the economic fallout of the crisis, with `small businesses' appearing prominently in both categories. Notably, China is referenced substantially in the {\tt InText} category but is more subdued in the {\tt InHashtag} case.
        \item Some generic terms such as `thank' appear in tweets about Covid - typically related to thanking caregivers and first responders. 
    \end{itemize}
On the issue of George Floyd's murder,
    \begin{itemize}
        \item A number of Democratic politicians connect George Floyd's murder to police brutality, with the term appearing significantly in both tweet text and in hashtags. Democrats also connect his death to the killings of Breonna Taylor and Ahmaud Arbery, suggesting a recognition of larger social patterns at play. Lastly, they  mention the party-sponsored Justice in Policing Act passed by the House as a response to calls for changes in policing.
        \item In contrast, while some Republicans addressed George Floyd's death, condemning it and calling for justice, there is little discussion of 'brutality' or talk about systemic problems with policing. Unlike Democrats, Republicans do not connect the Floyd case with other contemporary cases of violence against Black Americans. Further, looking at hashtags, we see that there are also frequent mentions of 'protest', 'right', and 'violence', which suggests a different tone to the one taken by Democrats.
        \item We have explored the use of therm `violence' by Republicans over time to assess how its use evolved.
    \end{itemize}
On the Black Lives Matter movement,
    \begin{itemize}
        \item Here, we see that Democrats tweet about police brutality, Breonna Taylor and George Floyd. However, we note that there are far less significant secondary words when looking at the Black Lives Matter tweets.
        \item Pulling Republican tweets with Black Lives Matter hashtags only yielded 6 tweets and is therefore not representative of a wider discourse. While there were 126 tweets in the {\tt InText} category, most were posted prior to the current wave of protests and relate to the Bureau of Land Management while only 21 were about the Black Lives Matter movement. This in itself is suggestive with regards to the involvement of Republicans in the recent discourse.
        \item We assessed the use of the terms `defund' and `police' in tweets after George Floyd's murder on May 25.
    \end{itemize}
\FloatBarrier

\section{Discussion}

The parties diverge substantially on all three issues of interest. Firstly, on COVID-19, we see that Democrats characterize the crisis through narratives emphasizing health care and public health, which is consistent with the party's recent, albeit tumultuous, shift towards prioritizing health care in their political platform. Republicans, on the other hand, highlight the effect of the shutdowns on the economy through an emphasis on small businesses, possibly building a narrative to justify re-opening, a recent political priority. This can be seen in context of the Trump administration's concerted efforts to push state and local governments to reopen businesses and schools. We also note that Republicans consistently mention China in their tweets about COVID-19, which is consistent with the party's recent anti-China rhetoric.

On the issue of George Floyd's murder, we observe that both parties mention his name alongside words like justice and police. However, while Democrats linked this murder to the killings of other Black individuals, and specifically named police brutality, we see that Republicans speak much more of protests and violence. Delving into the Republican discussion around `violence', we see that, in the two weeks immediately after the murder, Republicans posted 272 tweets that included the term `violence'. Seventy-three of these referred to Floyd by name. These tweets largely included condemnations of the murder, but would then pivot to criticizing perceived rioting, violence and looting. In the three subsequent weeks, there were 154 tweets about violence, but only six referred to Floyd by name. This indicates a shift in narrative, away from an emphasis on  Floyd's murder and towards a wider discussion of systemic racism and American heritage, including the defacing and removal of Confederate symbols and statues. 

With regards to Black Lives Matter, we see that there are a number of Democratic tweets containing the hashtag, but no significant co-occurrences .This may point to the lack of a clear agenda on how to address or respond to calls for anti-racist action.While there are insufficient Republican tweets to draw meaningful conclusions. This in itself is suggestive, indicating that the phrase in the hashtag \#BlackLivesMatter is not one with which Republicans want to associate. 

Finally, we  analyzed tweets with terms `defund' and `police' to examine the Republican position further. Of the 321 tweets posted after May 25 that contained these terms, 307 were posted by Republicans, either criticizing the growing `\#DefundThePolice' campaign, defending law enforcement, or criticizing Democratic opponents for supporting police defunding. The only tweets by Democrats that contained these terms appeared to distance themselves from the campaign - defending themselves from allegations by Republicans that they were in favor of it, or bluntly stating that they were against defunding the police. One exception was a tweet by Congresswoman Rashida Tlaib in full support, made significant by its rarity. Thus, we see that politicians at large, across both parties, did not show support for this campaign, despite its centrality in the anti-racist discourse.

\bibliographystyle{ACM-Reference-Format}
\bibliography{references}
\FloatBarrier











\end{document}